\documentclass[reprint]{revtex4-1}

\usepackage{graphicx}
\usepackage{dcolumn}

\begin{document}

\title{An Analysis of Physics Graduate Admission Data}

\author{Jan A. Makkinje}
\affiliation{Boston University \\ 590 Commonwealth Avenue, Boston MA, 02215}

\date{\today}

\begin{abstract}

We use self-reported data from 2011--2014 and determine the academic profile of accepted students at physics graduate programs in the United States. We analyze the accepted students' grade point averages and physics Graduate Record Examination (GRE) scores. We also compare the physics GRE scores of accepted students to their grade point averages.

\end{abstract}

\pacs{Valid PACS appear here}
\maketitle


\section{Introduction}

Many undergraduate students would like to continue their studies in physics and would prefer to be admitted to top ranked programs. However, often students have little information on the typical academic profile for students admitted to these programs.

During the graduate admissions process, prospective graduate students often discuss their prospects online of being admitted to various graduate programs. Many of these posts contain information about students' credentials such as grade point averages, physics Graduate Record Examination (GRE) score, and their general GRE scores. These posts also contain information about their status, such as whether they are domestic, international, or international students with a degree from an American University. By collecting information from these posts, we can gain insight into some trends that occur in admissions as well obtain information about acceptances based on the rank of the graduate programs.

\section{GPA}

\begin{figure}[h!]
\centering
\includegraphics[height=2.5in]{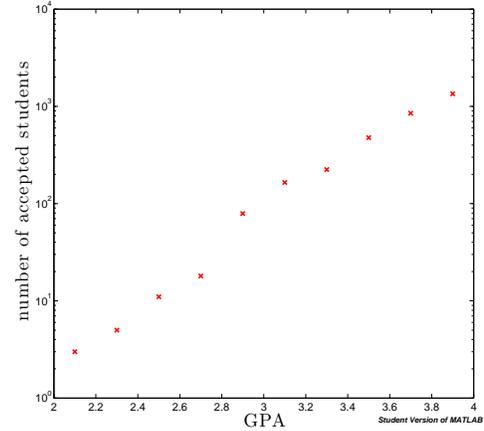}
\caption{Semilog plot of the number of accepted students as a function of their GPA. The number of acceptances increases with students' GPA. The slope is $\approx 2.5$.}
\label{fig:GPAAll}
\end{figure}

We first look at all the data available, not taking into consideration the international or domestic status of students. We normalize the grade point average (GPA) to a 4.0 scale so that GPAs can be compared. The number of accepted students with a given GPA is shown in Fig.~\ref{fig:GPAAll}. This data shows an exponential dependence of the number of accepted students as a function of their GPA.

\section{Physics GRE}

\begin{figure}[h!]
\centering
\includegraphics[height=2.5in]{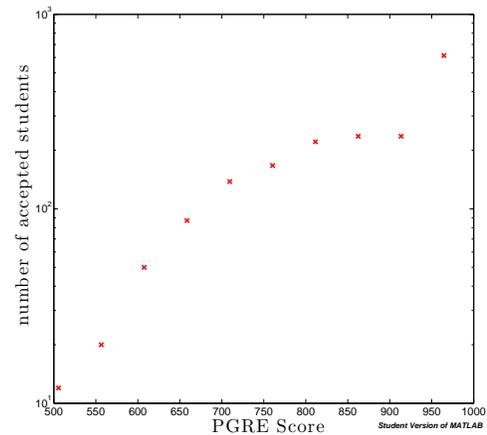}
\caption{Semilog plot of the total number of accepted students, regardless of their domestic or international status, as a function of their physics GRE score. Scores between $\approx{830}$ and $\approx{920}$ appear to be considered equally in graduate admissions.}
\label{fig:PGREAll}
\end{figure}

The physics GRE is an important quantity for graduate admissions and has a score between 200 and 990. Despite its shortfalls, the physics GRE allows graduate programs to compare students from different colleges on the same scale. The number of accepted students versus their physics GRE scores in is shown Fig.~\ref{fig:PGREAll}. This data shows a sigmoidal dependence for scores less than $\approx 920$ and then increases sharply. The results indicate that a higher physics GRE score is helpful, but increased scores greater than $\approx 830$ do not significantly increase the number of acceptances. However, near perfect scores are very advantageous in graduate admissions.

\begin{figure}[h!]
\centering
\includegraphics[height=2.5in]{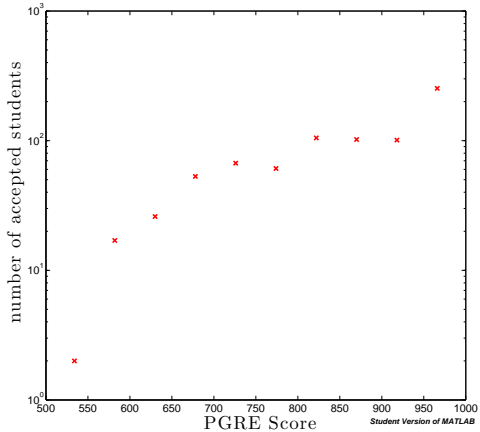}
\caption{Semilog plot of the total number of accepted international students as a function of their physics GRE scores. Note that the number of acceptances is less for lower scores.}
\label{fig:PGREInternational}
\end{figure}

Many universities outside of the United States use different grading scales and have different standards for students. As a result, the physics GRE is particularly important for evaluating international students. In Fig.~\ref{fig:PGREInternational} we show data for international students accepted to American universities. The plot shows the same overall trend, but increases in steps. These results may reflect minimum physics GRE score requirements at some universities. 

\section{Physics GRE and GPA}

\begin{figure}[h!]
\centering
\includegraphics[height=2.5in]{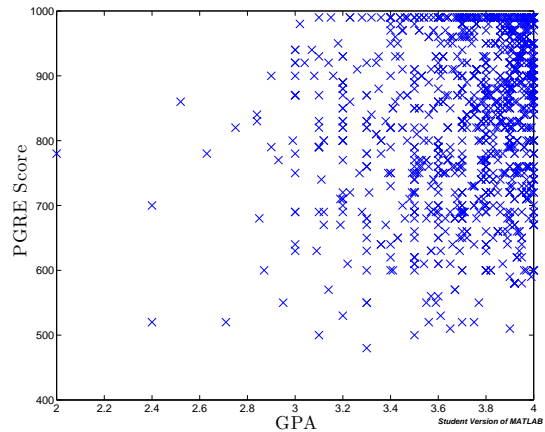}
\caption{Physics GRE scores versus GPA for admitted students. Note that a high GPA is more important than a high physics GRE.}
\label{fig:GPAGREAll}
\end{figure}

To obtain a better understanding of scores needed to obtain graduate admission, we look at the relation between the physics GRE and the GPA of admitted students. In Fig.~\ref{fig:GPAGREAll} we see that a higher GPA can generally compensate for a lower physics GRE score. That is, a wide range of admitted students' physics GRE scores correspond to students with very high GPAs. As expected, a high GPA and physics GRE scores leads to a greater number of acceptances.

\section{Highly ranked programs}

We rank the schools according to the U.S. News \cite{news} and calculate average data based on all of these schools. The average GPA of students accepted at a top ten ranked graduate program in 2011--2014 is $3.87 \pm 0.11$, compared to the average of $3.74 \pm 0.21$ for students admitted to all physics graduate programs. In addition, the average physics GRE score for top ten admitted students was found to be $902 \pm 77.5$. in comparison to an average score of $845.5 \pm 106.6$. As expected, highly ranked programs have higher average GPA and physics GRE scores, but also have lower standard deviations. 

\section{Conclusions}

It is important to note the shortcomings of this data set. The most important is that the data is self-reported. The sample size is limited and to some extent biased. We expect that most students who self-report this data are very interested in continuing to graduate programs and will tend to be better students. Moreover, in the analysis of all students this data does not correct for the quality of the programs to which students are applying. We also did not take into consideration research experience, personal statements, letter of references, and other important aspects that factor into graduate admission decisions. Finally, the reported GPAs are overall GPA's, not GPAs in physics courses. We expect students with a lower overall GPA and a high physics GPA to have more acceptances to graduate school than someone with a high overall GPA but a low physics GPA.

We hope that the data and our analysis will be of help prospective students gain a better understanding of where to apply based on their scores. In particular, the average GPA and physics GRE scores as well as the distribution could be helpful in assessing how competitive a student's application might be.

\section{Acknowledgements}

The author would like to thank Professor Harvey Gould for useful conversations and help in writing.
\\

\end{document}